\documentclass[11pt,twoside]{article}
\usepackage{asp2014}

\aspSuppressVolSlug
\resetcounters

\begin{document}

\title{Practical Provenance in Astronomy}

\author{Mathieu~Servillat,$^1$ Fran\c{c}ois Bonnarel,$^2$ Mireille Louys,$^{2,3}$ and Mich\`{e}le Sanguillon$^4$}
\affil{$^1$Laboratoire Univers et Th\'{e}ories, Observatoire de Paris, Université PSL, CNRS, Université de Paris, 92190 Meudon, France; \email{mathieu.servillat@obspm.fr}}
\affil{$^2$Centre de Donn\'{e}es astronomiques de Strasbourg, Observatoire Astronomique de Strasbourg, Universit\'{e} de Strasbourg, CNRS-UMR 7550, Strasbourg, France}
\affil{$^3$ICube Laboratory, Universit\'{e} de Strasbourg, CNRS-UMR 7357, Strasbourg, France}
\affil{$^4$Laboratoire Univers et Particules de Montpellier, Universit\'{e} de Montpellier, CNRS/IN2P3, France}

\paperauthor{Mathieu~Servillat}{mathieu.servillat@obspm.fr}{0000-0001-5443-4128}{LUTH - Observatoire de Paris}{}{Meudon}{}{92190}{France}
\paperauthor{Fran\c{c}ois Bonnarel}{francois.bonnarel@astro.unistra.fr}{ }{Universit\'{e} de Strasbourg, CNRS}{Observatoire astronomique de Strasbourg - UMR7550}{Strasbourg}{}{67000}{France}
\paperauthor{Mireille Louys}{mireille.louys@unistra.fr}{0000-0002-4334-1142}{Universit\'{e} de Strasbourg, CNRS} {ICube Laboratory - UMR7357}{Strasbourg}{}{67000}{France}
\paperauthor{Mich\`{e}le Sanguillon}{Michele.Sanguillon@umontpellier.fr}{0000-0003-0196-6301}{CNRS, Universit\'{e} de Montpellier}{LUPM}{Montpellier}{ }{34095}{France}




  
\begin{abstract}

Recently the International Virtual Observatory Alliance (IVOA) released a standard to structure provenance metadata, and several implementations are in development in order to capture, store, access and visualize the provenance of astronomy data products. This BoF will be focused on practical needs for provenance in astronomy. A growing number of projects express the requirement to propose FAIR data (Findable, Accessible, Interoperable and Reusable) and thus manage provenance information to ensure the quality, reliability and trustworthiness of this data. The concepts are in place, but now, applied specifications and practical tools are needed to answer concrete use cases. 
During this session we discussed which strategies are considered by projects (observatories or data providers) to capture provenance in their context and how a end-user might query the provenance information to enhance her/his data selection and retrieval. The objective was to identify the development of tools and formats now needed to make provenance more practical needed to increase provenance take-up in the astronomical domain. 

\end{abstract}

\section{The path to Open Science}

Open science is the movement to make scientific research (including publications, data, physical samples, and software) and its dissemination accessible to all levels of an inquiring society, amateur or professional.

In this context, the European Open Science Cloud (EOSC) is an environment for hosting and processing research data to support science. The aim is to make research data interoperable and machine actionable following the FAIR principles\footnote{\url{https://www.go-fair.org/fair-principles}} \citep{Wilkinson2016}, by federating existing research data infrastructures in Europe and realising a web of FAIR data and related services for science, .

Within EOSC, the ESCAPE project\footnote{\url{https://projectescape.eu}} (European Science Cluster of Astronomy \& Particle physics ESFRI research infrastructures) brings together the astronomy, astroparticle and particle physics communities and puts together a cluster with ESFRI projects with aligned challenges of data-driven research. Keeping and exposing provenance information is one of those challenges.

A workshop\footnote{\url{https://indico.in2p3.fr/event/21913/page/2641-summary}} dedicated to the management of provenance took place in September 2020 within ESCAPE, in line with this ADASS session on practical provenance. The discussions led to a clarification of the requirement for structured provenance (\S\ref{req_struct_prov}), a more precise terminology (\S\ref{prov_termino}), and the identification of key concepts associated with the management of provenance information (\S\ref{prov_key_concepts}).

\section{The IVOA Provenance Data Model}

The IVOA has released a recommendation for a Provenance Data Model \citep{2020ivoa.spec.0411S}. The core model and definition of provenance is as defined by the W3C \citep{std:W3CProvDM}: information about entities, activities, and people (agents) involved in producing a piece of data or thing, which can be used to form assessments about its quality, reliability or trustworthiness.

Provenance is related by definition to the origin of a product (where does it come from?), but also the path followed to generate this product (what has been done?). 
Provenance is thus seen as a chain of activities and entities, used and generated. With the core data model, the basic objectives are achieved: use of unique identifiers, traceability of the operations, connection with contacts for further information, citation or acknowledgement. By following the full IVOA data model, more advanced questions are answered: What happened during each activity? How was the activity tuned to be executed properly? What kind of content is in the entities? 

The data model is a base for the development of tools and services, see e.g.: \citet{P9-89_adassxxx, P9-250_adassxxx, P9-216_adassxxx, 2020ASPC..522..199S, 2020ASPC..522..545S}.

\section{Requirement for structured provenance}
\label{req_struct_prov}

We often realize too late that there are missing elements or links in the provenance. The capture of the provenance should thus be structured, as detailed as possible, and as naive as possible (simply record what happens).

There are clear advantages to retain this information as structured, machine-readable data, in particular in the context of Open Science:
    
\begin{itemize}
\setlength{\itemsep}{1mm}
\setlength{\parskip}{0pt}
    \item Explicitly required through the \textbf{FAIR principles} that indicate to include \textbf{rich metadata}, following standard data model, protocols and formats, with \textbf{detailed provenance}.
    \item \textbf{Quality / Reliability / Trustworthiness} of the products: the simple fact of being able to show its provenance is sufficient to give more value to a product, and if the provenance information is detailed, the value will be higher.
    \item \textbf{Reproducibility requirement} in many projects: provenance details are essential to be able to rerun each activity (maybe testing and improving each step); Having this information, it may not be necessary to keep every intermediate file that is easily reproducible (hence a possible gain on storage space and costs).
    \item \textbf{Debugging}: with detailed provenance, it is not necessary to restart from scratch, as one can locate in the provenance graph the faulty parts or the products to be discarded, and reprocess only from the identified failing steps.
\end{itemize}

\section{Some terminology}
\label{prov_termino}

The word "provenance" is used to refer to different aspects depending on the persons or the goals involved. Provenance may be used for internal data management, or to improve the scientific exploitation of a data product. It may be stored inside the data file or separately (external file or database). We propose here a base for the definition of provenance categories.
\begin{itemize}
\setlength{\itemsep}{1mm}
\setlength{\parskip}{0pt}
  \item \textbf{full provenance}: graph/tree/chain of activities and entities up to the raw data. This information is not hosted by the entities themselves, but stored on an external server, or as separate files.
  \item \textbf{minimum provenance}: information attached to an entity as a list of keywords that gives some context and info on last activity (general process/workflow, software versions, contacts...), maybe including the list of  used entities, so that a full provenance may be reconstructed from  minimum provenance, but such information is not always preserved.
  \item \textbf{end-user/specific "provenance"}: information attached to an entity, as a list of keywords or data that provides key information to use/analyse the entity (e.g. for CTA: event class, event type, telescope configuration, sky conditions, reconstruction method,...). This information may be extracted from the full provenance, or inversely, used to enrich a reconstructed provenance.
\end{itemize}

\section{Applying the model}
\label{prov_key_concepts}

Different concepts were identified concerning provenance in various use cases. We distinguish \textbf{on-top} provenance handling, where data products/collections already exist and is used to reconstruct a provenance graph, and \textbf{inside} provenance processing, where the provenance information is saved during the execution of the processing activities.

The use of \textbf{unique identifiers} is recommended and implies to think about provenance capture at the conception phase of a project. One also has to evaluate in advance the necessary and sufficient \textbf{granularity} (what steps? what objects?) and \textbf{level of details} (inclusion of descriptions and configuration details?).

The challenges in provenance management encompass the \textbf{capture}, \textbf{storage}, \textbf{access}, and \textbf{visualization} of the provenance information.

\section{Discussion topics}

The discussion was organised around several topics, following the priorities of the participants as reported in a preliminary questionnaire. Those topics will be discussed during forthcoming ESCAPE events, in collaboration with IVOA working groups:  

\textbf{1.} Defining the content of a minimum provenance, with a list of keywords related to the last activity and context.

\textbf{2.} Serializing provenance, both in a human- and machine-readable way.

\textbf{3.} Provenance and workflows, with workflow information simply attached to provenance (as used entities), or mapped to the model.

\textbf{4.} Evolving from provenance "on-top" to provenance "inside", e.g. with generic tools to capture provenance at the execution.

\textbf{5.} Provenance storage in a database and related interfaces.

\textbf{6.} Provenance exploration and visualization, based on access protocols (ProvTAP, see \citealt{2019ASPC..523..313B}, or ProvSAP, see \citealt{P9-89_adassxxx}.), and tools such as the \texttt{voprov} Python package.


\acknowledgements We acknowledges support from the ESCAPE project funded by the EU Horizon 2020 research and innovation program (Grant Agreement n.824064). Additional funding was provided by the INSU (Action Sp\'ecifique Observatoire Virtuel, ASOV), the Action F\'ed\'eratrice CTA at the Observatoire de Paris and the Paris Astronomical Data Centre (PADC).

\bibliography{B9-56}


\end{document}